# Stability analysis of DC microgrids with constant power load under distributed control methods

Zhangjie Liu[a], Mei Su[a], Yao Sun[*a], Hua Han[a], Xiaochao Hou[a], Josep M. Guerrero[b]

[a]*School of Information Science and Engineering, Central South University, Changsha, 410083, China*
[b]*Department of Energy Technology, Aalborg University, Aalborg East, DK-9220, Denmark*
* yaosuncsu@gmail.com

**Abstract**

Constant power loads (CPLs) often cause instability due to its negative impedance characteristics. In this study, the stability of a DC microgrid with CPLs under a distributed control that aims at current sharing and voltage recovery is analyzed. The effect of the negative impedance on the behavior of distributed controller are investigated. The small-signal model is established to predict the system qualitative behavior around equilibrium. The stability conditions of the system with time delay are derived based on the equivalent linearized model. Additionally, eigenvalue analysis based on inertia theorem provides analytical sufficient conditions as a function of the system parameters, and thus it leads to a design guideline to build reliable microgrids. Simulations are performed to confirm the effectiveness and validity of the proposed method.

## 1. Introduction

Driven by environmental concerns, renewable energy sources, such as photovoltaics and wind generation, are being rapidly deployed (Simpson-Porco, Dörfler & Bullo, 2013; George, Zhong, et al., 2015; Schiffer et al., 2016; Bidram1 et al., 2014; Song et al., 2017). Microgrids have been identified as key components of modern electrical systems for facilitating the integration of renewable distributed generation units (Schiffer et al., 2014; Chang & Zhang, 2016). Microgrids can be divided into two types: alternating-current (AC) and direct-current (DC) microgrid (Sun et al., 2017). Recently, DC microgrids have attracted increasing attentions owing to their advantages, including their reliability and efficiency, simple control, robustness and natural interface for renewable source (Kakigano et al., 2010). As a result, DC microgrid consisting of multiple converters is increasingly used in applications such as aircrafts, space crafts, electric vehicles (Maknouninejad, et al., 2014).).

Usually, the main control objectives of a DC microgrid include sharing current, regulating voltage and maintaining stability (Han et al., 2017). There are mainly three methods for current sharing and voltage regulation: decentralized control, centralized control and distributed control.

The traditional V-I droop is a typical decentralized control which is convenient, inexpensive and efficient, however, with the drawbacks: voltage sag and biased power sharing (Augustine et al., 2015). For this situation, several improved decentralized methods are proposed (Huang et al., 2015). Although the performance has improved, the shortcomings are not completely overcome.

The centralized control method has been widely used in DC microgrids. All distributed generators (DGs) can realize the targets of current sharing and voltage recovery via commands from the central controller unit, which collects the global information (Guo et al., 2014). Despite its satisfactory performance, centralized control requires complex communication networks and is thus vulnerable to link failures.

To overcome these drawbacks, distributed control strategies have been proposed. The key feature of the distributed control method is the consensus algorithm, which just needs the neighbor information (Nasirian et al., 2015; Nasirian et al., 2014; Moayedi et al., 2016; Shafiee et al., 2014; Zhao & Dörfler, 2015). Thus, highly accurate current sharing and voltage regulation can be realized via a sparse communication network (Behjati et al., 2014; Meng et al., 2016).

However，the DC microgrid with CPL tends to be unstable when traditional decentralized control or distributed control is implemented independently. Stability issues of the DC microgrid with CPL under decentralized control have been investigated. In order to realize current sharing, the small-signal stability of a system with CPLs under droop control has been analyzed in (Sandeep & Fernandes, 2013; Tahim et al., 2015; Su, Liu, Sun, Han & Hou, 2016). A reduced-order linearized model is derived, wherein the transient process is ignored (Sandeep & Fernandes, 2013; Tahim et al., 2015). These studies show that if the droop coefficient is larger than the equivalent negative impedance of the CPL, the system would be stable. A high-dimensional model has been proposed for analyzing the transient processes of the converter (Su, Liu, Sun, Han & Hou, 2016). By using the quadratic eigenvalue problem theories (Tisseur & Meerbergen, 2001), the stability of the linearized system is analyzed, and a wide stability solution is obtained.

With the development of the communication technology, distributed control becomes more and more popular in DC microgrid. Nevertheless, stability of the system with CPL under distributed control has never been studied concretely. What's worse, possible time delay would further deteriorate the stability problem. So, it is vital to develop the stability analysis of the system under distributed control.

In this paper, we propose a distributed control method that not only overcomes the instability of the CPL, but also realizes current sharing and voltage regulation. This method can be treated as a combination of V-I droop (inherently virtual resistance) and distributed control. Moreover, the stability condition of the system with time delay is also obtained. The main contributions of this study are summarized as follows.

- A distributed control method is proposed that not only overcomes the instability of the CPL, but also realizes current sharing and voltage regulation.
- The small-signal stability of the system with a CPL under distributed control is analyzed, and the analytic sufficient conditions are obtained. The relation among the line resistances, control parameters, reference voltage and the maximum load that keep the system stable is obtained.
- This paper provides an efficient method for stability analysis of two typical matrices (see Theorem 1 and 3). These problems can be effectively solved using the proposed method.
- The stability condition of the system with time delay is obtained, and the validity of the proposed method is tested by simulations.

The paper is organized as follows. Section 2 introduces the distributed control framework. The stability analysis and the sufficient conditions are introduced in Section 3. The simulation results are presented in Section 4. The conclusions are drawn in Section 5.

## 2. Distributed control framework

### 2.1. Basic model and assumptions

The DC microgrid, which comprises multiple parallel DC/DC converters with a CPL, is shown in Fig.1.(a). It contains a physical network and a communication network and is modeled according to the following assumptions:
1) The response of the buck converter is sufficiently fast that dynamics can be neglected. That is, the DC/DC converters can be treated as ideal controllable voltage sources.
2) The loads are ideal CPLs. In fact, because the response of the output regulating controllers of the point of load (POL) converters is fast enough, all the POL converters attached to the load could be regarded as CPLs (Su, Liu, Sun, Han & Hou, 2016).
3) The resistance of the common bus is zero; hence, all loads are regarded as one common CPL.
4) The cable is purely resistive. In low-voltage DC microgrid, the cable inductance can be neglected.

For constant power loads, the power balance equation should be satisfied.

$$\begin{cases} u_L \sum_{i=1}^{n} i_i = P \\ u_i = i_i r_i + u_L \end{cases} \quad (1)$$

where $u_L$ represents the voltage of the DC bus, $P$ is the power of the load, and $r_i$ represents the resistance of the cable between the $i^{th}$ DG to loads.

### 2.2. Graph theory

Fig.1.(b) shows the mapping of a cyber network to a physical DC microgrid. The nodes represent converters, and the edges represent the communication links for data exchange. In distributed control, all agents exchanges information only with their neighbors.

A graph is usually represented as a set of nodes $V_G=\{v_1, v_2,…,v_n\}$ connected by a set of edges $E_G \subset V_G \times V_G$, along with an associated adjacency matrix $A_G =[a_{ij}]\in R^{n\times n}$. $n$ is the number of nodes. The elements of $A_G$ represent the communication weights, where $a_{ij} > 0$ if the edge $(v_j, v_i)\in E_G$; otherwise, $a_{ij} =0$. Here, the matrix $A_G$ is assumed to be time-invariant. The in-degree matrix $D_G = diag\{d_i\}$ is a diagonal matrix with $d_i = \sum a_{ij}$. The Laplacian matrix is defined as $L = D_G - A_G$, and its eigenvalues determine the global dynamics. For a connected graph, there is at least one spanning tree, and $\ker(L) = \text{span}(1_n)$, where $1_n = [1\ 1\ …\ 1]^T$ (Olfati-Saber & Murray, 2004).

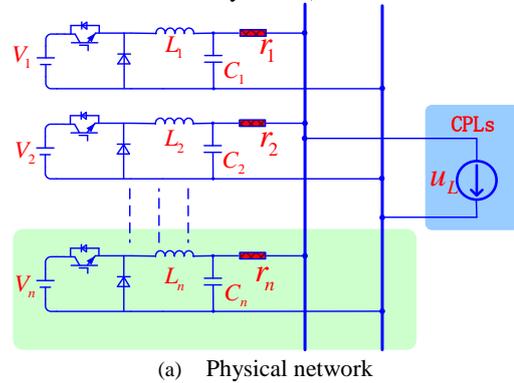

(a) Physical network

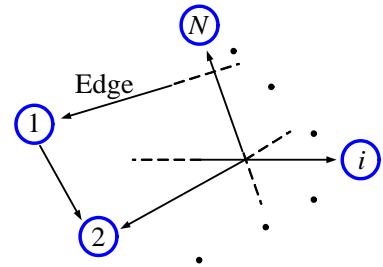

(b) Communication network
Fig.1. The structure of a DC microgrid

### 2.3. Stabilizing distributed control

To realize proportional current sharing and excellent load voltage regulation, a distributed control method is proposed. To overcome the instability of the CPL, stabilization measures are necessary. Because damping can mitigate oscillation, virtual resistances are employed for auxiliary stabilization and improving the transient performance. The control diagram is illustrated in Fig. 2. The output voltage for each converter can be expressed as

$$u_i = v_{ref} + \delta i_i + \delta u_i - c_i i_i \quad (2)$$

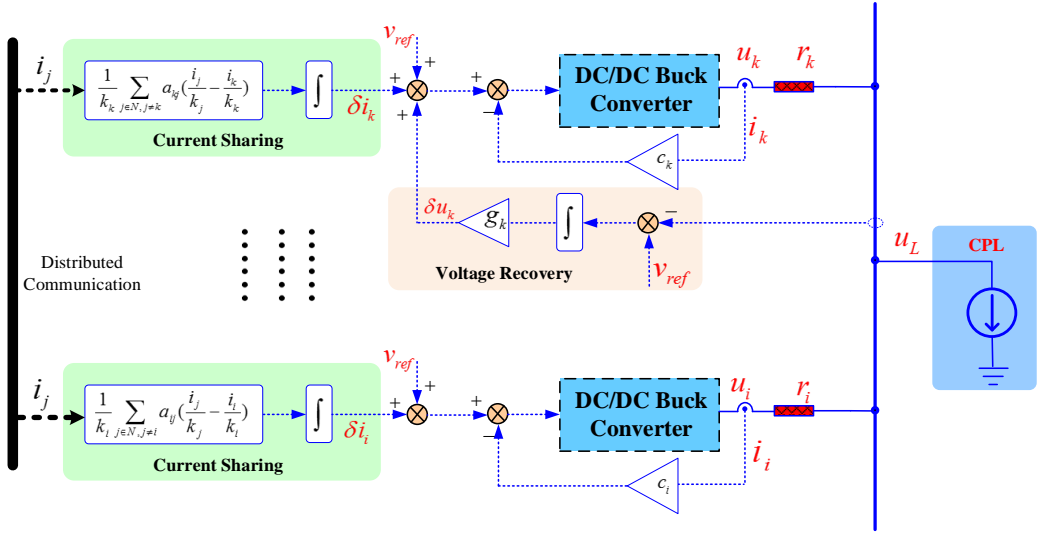

Fig. 2. System control scheme. Only a small number of DGs is needed to sample the load voltage to generate feedback.

where $u_i$, $i_i$, $\delta i_i$, and $\delta u_i$ represent the output voltage, output current, current-correction terms and the voltage-correction terms, respectively, for the $i^{th}$ converter. $v_{ref}$ represents the rated point of the DC bus voltage; and $c_i$ represents the virtual resistance (or droop gain of V-I droop).

The current correction term is designed as follows:

$$\delta i_i = \frac{b_1}{k_i}\int \sum a_{ij}\left(\frac{i_j}{k_j}-\frac{i_i}{k_i}\right)dt \quad (3)$$

where $k_i$ ($k_i>0$) is the current-sharing proportionality coefficient, $b_1$ is a positive gain coefficient, and $a_{ij}$ represents the communication weight. If there is a communication link between nodes $i$ and $j$, $a_{ij}= a_{ji}>0$; otherwise, $a_{ij}=0$. Rewriting (3) in matrix form yields

$$\delta i = -b_1\int KLKi\,dt \quad (4)$$

where $\delta i = [\delta i_1 \quad \delta i_2 \quad \cdots \quad \delta i_n]^T$, $K = diag\{1/k_i\}$, and $L$ is the Laplacian matrix of the communication graph. Usually, the communication network is sparse and has at least one spanning tree. As indicated by (4), one of the advantages of this design is that it maintains the symmetry of the system.

To recover the load voltage directly, the voltage regulator can be expressed as

$$\delta u_i = b_2 \int g_i\left(v_{ref}-u_L\right)dt \quad (5)$$

where $g_i$ is the weight coefficient of the $i^{th}$ converter; $b_2$ ($b_2 > 0$) is the auxiliary gain coefficient; and $u_L$ is the load voltage. $g_i >0$ if there is a communication link between the DG and the DC bus; otherwise, $g_i=0$. Typically, to enhance the reliability of the system, two or three DGs must participate in the voltage recovery. This can significantly reduce the communication cost while maintaining high reliability. By rewriting (5) in matrix form, we obtain the following:

$$\delta u = b_2 \int \left(v_{ref}-u_L\right)g\,dt \quad (6)$$

where $g = [g_1 \quad g_2 \quad \cdots \quad g_n]^T$.

By combining (3) and (5) with (2), the voltage reference of the DC/DC converter is determined.

### 2.4. Steady-state analysis

According to (2), (4) and (6), the voltage is expressed as

$$u = v_{ref}1_n - Ci + \int -b_1KLKi + b_2\left(v_{ref}-u_L\right)g\,dt \quad (7)$$

where $u=[u_1\ u_2\ ...\ u_n]^T$ and $C=diag\{c_i\}$. In the steady state, this yields

$$-b_1KLKi + b_2\left(v_{ref}-u_L\right)g = 0_n \quad (8)$$

where $0_n$ is an $n$-dimension vector with full of zeros.

**Corollary 1**. According to (8), equations (9)-(10) hold if $\sum k_i g_i \neq 0$:

$$\frac{i_1}{k_1} = \frac{i_2}{k_2} = \cdots = \frac{i_n}{k_n} \quad (9)$$

$$u_L = v_{ref} \quad (10)$$

**Proof**. Because $K$ is a full rank matrix, there is a matrix $K^{-1}$ that satisfies

$$b_1LKi + b_2\left(v_{ref}-u_L\right)K^{-1}g = 0_n \quad (11)$$

As $L$ is a balanced symmetric Laplacian matrix, $\ker(L) = span(1_n)$. That is, $1_n^T L = 0_n^T$. Thus, the following can be derived from (11)

$$b_1 1_n^T LKi + b_2\left(v_{ref}-u_L\right)1_n^T K^{-1}g = 0 \quad (12)$$

Because $1_n^T L = 0_n^T$, and $1_n^T LKi = 0$, (10) obviously holds if $\sum k_i g_i \neq 0$. By substituting (10) into (12), we obtain

$$LKi = 0_n \quad (13)$$

It is easy to find that $Ki \in span(1_n)$, which is equivalent to (9). Thus, the proof is accomplished.

**Remark:** As **Corollary** 1 shows, voltage recovery does not require all DGs to take part in. That is, as long as there is more than one DG to communicate with the load, the voltage recovery can be realized. Then, several DGs in close proximity can be used to collect the load voltage, reducing the communication cost.

Note that $u^*$ and $i^*$ are the output voltage and current vector of the DGs, respectively, and $u_L^*$ is the voltage of the DC bus when the system is in the steady state. Because $u^*$ and $i^*$ satisfy (1), (9), and (10), we obtain

$$\begin{cases} i_i^* = \dfrac{k_i}{\sum_{j=1}^{n} k_j} \dfrac{P}{u_L^*} \\ u_i^* = u_L^* + \dfrac{r_i k_i}{\sum_{j=1}^{n} k_j} \dfrac{P}{u_L^*} \end{cases} \quad (14)$$

## 3. Stability analysis

The next parts employ the following five lemmas:

**Definition 1**. If $A \in \mathrm{R}^{n \times n}$, define:

$i_+(A)$ = the number of eigenvalues of matrix $A$, counting multiplicities, with a positive real part;

$i_-(A)$ = the number of eigenvalues of matrix $A$, counting multiplicities, with a negative real part;

$i_0(A)$ = the number of eigenvalues of matrix $A$, counting multiplicities, with a zero real part.

Then, $i_+(A) + i_-(A) + i_0(A) = n$. The row vector is given

$$i(A) = [i_+(A), i_-(A), i_0(A)] \quad (15)$$

Which is called the inertia of matrix $A$ (Horn & Johnson, 1986).

**Definition 2**. A matrix $A \in \mathrm{R}^{n \times n}$ is said to be positive stable if $i_+(A) = n$; $A$ is semi-positive stable if $i_-(A) = 0$.

**Lemma 1.** Let $A$ be a positive semidefinite matrix, and let $H$ be a Hermitian matrix. Then

$$i_+(AH) \leq i_+(H), i_-(AH) \leq i_-(H) \quad (16)$$

(The proof is presented in (Ostrowski & Schneider, 1962).)

**Lemma 2.** $M, K \in \mathrm{R}^{n \times n}$. If $M$ is a real symmetric positive semidefinite matrix and $K$ is a real symmetric matrix, all the eigenvalues of $MK$ ($KM$) are real (Hong & Horn, 1991).

**Lemma 3**. Schur complements and determinantal formulae. We let $A$ be a square matrix and partition $A$ as

$$A = \begin{bmatrix} A_{11} & A_{12} \\ A_{21} & A_{22} \end{bmatrix} \quad (17)$$

If $A_{11}$ and $A_{22}$ are both nonsingular, we obtain

$$\begin{cases} \det(A) = \det(A_{11})\det(A_{22} - A_{21}A_{11}^{-1}A_{12}) \\ \det(A) = \det(A_{22})\det(A_{11} - A_{12}A_{22}^{-1}A_{21}) \end{cases} \quad (18)$$

(The proof is presented in (Horn, & Johnson, 2012).)

**Lemma 4**. If $\lambda_1 \leq \lambda_2 \leq \cdots \leq \lambda_n$ are the eigenvalues of a real symmetric matrix $A$, and $\beta_1 \leq \beta_2 \leq \cdots \leq \beta_n$ are the eigenvalues of a real symmetric matrix $B$, we obtain

$$\lambda_i + \beta_1 \leq \eta_i \leq \lambda_i + \beta_n \quad (19)$$

where $\eta_1 \leq \eta_2 \leq \cdots \leq \eta_n$ are the eigenvalues of matrix $A+B$.

(The proof is presented in (Meyer, 2000).)

**Lemma 5**. Define $\Lambda = c_1 \oplus c_2 \cdots \oplus c_{n-1} \oplus 0, M_1 = \Lambda + ab^T$, $M_2 = M_1 + ba^T, a = [a_1 \; a_2 \; \cdots \; a_n]^T$, and $b = [b_1 \; b_2 \; \cdots \; b_n]^T$ Then,

$$\begin{cases} \det(M_1) = a_n b_n \prod_{i=1}^{n-1} c_i \\ \det(M_2) = \prod_{i=1}^{n-1} c_i \left( 2 a_n b_n - \sum_{i=1}^{n-1} \dfrac{1}{c_i}(a_n b_i - a_i b_n)^2 \right) \end{cases} \quad (20)$$

(The proof is presented in the Appendix of this paper.)

**Lemma 6**. All the zeros of the function of $g(z)=z+\beta e^{-z}$, $\beta= be^{i\alpha}$, $(b > 0, |\alpha| < \pi)$, have negative real part if and only if

$$\dfrac{\pi}{2} - b > |\alpha| \quad (21)$$

(The proof is in (DeFazio & Muldoon, 2005).

### 3.1. Small signal approximate model

As shown in (9), (10) and (14), if the system is stable, current sharing and voltage recovery can be easily accomplished. However, maintaining the stability of the system under CPLs is challenging.

By differentiating (2), the system dynamic model can be obtained

$$\begin{cases} \dfrac{du_i}{dt} = -c_i \dfrac{di_i}{dt} + \dfrac{1}{k_i}\sum_{j=1}^{n} a_{ij}\left( \dfrac{i_j}{k_j} - \dfrac{i_i}{k_i} \right) + g_i \left( v_{ref} - u_L \right) \\ u_i = i_i r_i + u_L; \quad \left( \sum_{i=1}^{n} i_i \right) u_L = P \end{cases} \quad (22)$$

Clearly, it is a nonlinear differential-algebra equation. According to the Hartman–Grobman theorem, the behavior of a dynamical system in a domain near the hyperbolic equilibrium point is qualitatively the same as the behavior of its linearization near this point. Substituting $u_i = u_i^* + \Delta u_i$ and $i_i = i_i^* + \Delta i, u_L = u_L^* + \Delta u_L$ into (22), the linear approximation of the nonlinear system is obtained

$$\begin{cases} \dfrac{d\Delta u_i}{dt} = -c_i \dfrac{d\Delta i_i}{dt} + \dfrac{1}{k_i}\sum_{j=1}^{n} a_{ij}\left( \dfrac{\Delta i_j}{k_j} - \dfrac{\Delta i_i}{k_i} \right) - g_i \Delta u_L \\ \Delta u_i = \Delta i_i r_i + \Delta u_L \\ \sum_{i=1}^{n} \Delta i_i = -\dfrac{P}{v_{ref}^2} \Delta u_L \end{cases} \quad (23)$$

According to (23), the following can be derived

$$\Delta u = Z \Delta i \quad (24)$$

where

$$\begin{cases} \Delta i = [\Delta i_1 \; \Delta i_2 \; \cdots \; \Delta i_n]^T \\ \Delta u = [\Delta u_1 \; \Delta u_2 \; \cdots \; \Delta u_n]^T \end{cases} \quad (25)$$

and $Z = diag\{r_i\} - \frac{v_{ref}^2}{P}1_n1_n^T$. Note that $r_L = -\frac{v_{ref}^2}{P}$; thus, $r_L$ is the equivalent resistance of the CPL. By substituting (24) into (23), we derive the normal state equation:

$$\frac{d\Delta i_i}{dt} = -(C+Z)^{-1}(KLK+G)\Delta i_i \quad (26)$$

where $G = r_L g 1_n^T$. Then, the Jacobian matrix of the system is obtained as

$$J = -(C+Z)^{-1}(b_1 KLK + b_2 G) \quad (27)$$

*3.2. Stability conditions without communication delay*

Firstly, we assume that the system involves no time delay. Define $J_1 = (C+Z)^{-1}(b_1 KLK + b_2 G)$, if the real part of every eigenvalue of matrix $J_1$ is positive, the system is stable. According to the corollaries in (Su, Liu, Sun, Han & Hou, 2016), the matrix $Z$ has one negative eigenvalue and $n-1$ positive eigenvalues. Since the load resistance is far larger than the cable resistance and the virtual resistance, then $r_L+\max\{r_i+c_i\}<0$. Similarly, the matrices $C+Z$ and $(C+Z)^{-1}$ have one negative eigenvalue and $n-1$ positive eigenvalues. Thus, the problem is difficult. If $A$ is symmetric positive definite and $B$ is general positive definite, it is easily shown that the matrix $AB$ is positive stable and the matrix $A^{-1}$ is one of the Lyapunov solutions. Unfortunately, the matrix $(C+Z)^{-1}$ is not positive definite; thus, it is difficult to use the same method to analyze the inertia of this matrix.

In the matrix $J_1$, $Z$ and $L$ depend on the physical topology and communication topology, respectively, of the DC microgrid, $K$ and $C$ consist of specific control parameters. We assume that only $b_1$ and $b_2$ can be selected and designed to stabilize the system. Thus, the following questions arise:

1) Is there a pair of $b_1$ and $b_2$ that makes the matrix $J_1$ positive stable?
2) How to select the parameters to enhance the stability?

In this study, these problems are analyzed. To obtain the stability conditions, the work is divided in the next three steps.

We define $J_2 = (C+Z)^{-1}KLK, J_3 = (C+Z)^{-1}G$. Then $J_1 = b_1 J_2 + b_2 J_3$.

**Theorem 1.** The matrix $J_2$ is semi-positive stable if and only if

$$P < \frac{v_{ref}^2 \left(\sum_{i=1}^n k_i\right)^2}{\sum_{i=1}^n k_i^2 (r_i + c_i)} \quad (28)$$

**Proof.** According to **Lemma 2**, the eigenvalues of the matrix $J_2$ are all real. We define $\mu_1, \mu_2, \cdots, \mu_n$ as the eigenvalues of matrix $J_2$. There is a zero root because $L$ is a singular matrix; thus, we assume $\mu_1 = 0$ and $\mu_2 \leq \cdots \leq \mu_n$. To realize global synchronization, the communication graph must have at least one spanning tree, that is, $i(L) = [n-1 \ 0 \ 1]$. Because the matrix $K$ is nonsingular and symmetric, according to Sylvester's law of inertia, $i(KLK) = i(L)$. According to **Lemma 1**, we obtain the following

$$i_-(J_2) \leq i_-\left((C+Z)^{-1}\right) = 1 \quad (29)$$

That is, matrix $J_2$ has at most one eigenvalue with a negative real part. Thus, we can obtain the critical condition: matrix $J_2$ does not have an eigenvalue with a negative real part if and only if $\mu_2 \mu_3 \cdots \mu_n > 0$. Next, we investigate the relationship between the product of nonzero eigenvalues of matrix $J_2$ and the determinant of matrices $C+Z$, $L$ and $K$.

The characteristic equation of $J_2$ can be expressed as

$$\left|\lambda I - (Z+C)^{-1} KLK\right| = 0 \quad (30)$$

By multiplying through by $(C+Z)$, we obtain

$$\left|\lambda(C+Z) - KLK\right| = 0 \quad (31)$$

Because $K$ is nonsingular, (31) is equivalent to

$$\left|\lambda K^{-1}(C+Z)K^{-1} - L\right| = 0 \quad (32)$$

We define matrix $N$ as follows

$$N = \begin{bmatrix} I_{n-1} & 0_{n-1} \\ 1_{n-1}^T & 1 \end{bmatrix}$$

Obviously, $N$ is nonsingular, thus, (31) is equivalent to

$$\left|\lambda NK^{-1}(C+Z)K^{-1} - NL\right| = 0 \quad (33)$$

Note that $Q = K^{-1}(C+Z)K^{-1}$, in fact, the matrix $Q$ is contract with the matrix $(C+Z)$. So they share the same inertia, that is

$$i(Q) = i(C+Z) = [n-1 \ 1 \ 0] \quad (34)$$

Partition $Z_1$ and $L$ as

$$Q = \begin{bmatrix} Q_1 & \alpha \\ \alpha^T & q \end{bmatrix}, L = \begin{bmatrix} L_1 & \beta \\ \beta^T & d_n \end{bmatrix}$$

where $q$ is a positive scalar and $\alpha, \beta$ is a positive column vector. Then, (32) is equivalent to

$$\begin{vmatrix} \lambda Q_1 - L_1 & \lambda \alpha - \beta \\ \lambda\left(1_{n-1}^T Q_1 + \alpha^T\right) & \lambda\left(1_{n-1}^T \alpha + q\right) \end{vmatrix} = 0 \quad (35)$$

According to the law of the determinant, (33) is equivalent to

$$\lambda \begin{vmatrix} \lambda Q_1 - L_1 & \lambda \alpha - \beta \\ 1_{n-1}^T Q_1 + \alpha^T & 1_{n-1}^T \alpha + q \end{vmatrix} = 0 \quad (36)$$

If $1_{n-1}^T \alpha + q \neq 0$, according to **Lemma 3**, (34) is equivalent to

$$\lambda \left|\lambda Q_1 - L_1 - \frac{1}{1_{n-1}^T \alpha + q}(\lambda \alpha - \beta)\left(1_{n-1}^T Q_1 + \alpha^T\right)\right| = 0 \quad (37)$$

Simplifying (35) yields

$$\lambda\left|\lambda\left(Q_1 - \frac{1}{\left(1_{n-1}^T\alpha+q\right)}\alpha\left(1_{n-1}^T Q_1+\alpha^T\right)\right) - \left(L_1 - \frac{1}{\left(1_{n-1}^T\alpha+q\right)}\beta\left(1_{n-1}^T Q_1+\alpha^T\right)\right)\right| = 0 \quad (38)$$

For convenience, note that

$$Q_2 = Q_1 - \frac{1}{\left(1_{n-1}^T\alpha+q\right)}\alpha\left(1_{n-1}^T Q_1+\alpha^T\right)$$

$$L_2 = L_1 - \frac{1}{\left(1_{n-1}^T\alpha+q\right)}\beta\left(1_{n-1}^T Q_1+\alpha^T\right)$$

If $Q_2$ is nonsingular, (36) is equivalent to

$$\lambda\left|\lambda I - Q_2^{-1}L_2\right| = 0 \quad (39)$$

Clearly, (37) has a zero root, and the others are $\mu_2, \mu_3, \cdots, \mu_n$. Thus, we have

$$\mu_2\mu_3\cdots\mu_n = \det\left(Q_2^{-1}L_2\right) = \det\left(Q_2^{-1}\right)\det\left(L_2\right) \quad (40)$$

Then, if $\det(Q_2^{-1}L_2)$ is positive, the matrix $J_2$ is semi-positive stable. Furthermor, according to **Lemma 3**, we have

$$\det\left(Q_2^{-1}\right) = \frac{1}{\det(Q_2)} = \frac{1_{n-1}^T\alpha+q}{\det(Q)\det(N)} = \frac{1_{n-1}^T\alpha+q}{\det(Q)} \quad (41)$$

$$\det\left(L_2\right) = \frac{1_n^T Q 1_n}{1_{n-1}^T\alpha+q}\det\left(L_1\right) \quad (42)$$

According to (34), $\det(Q) < 0$. Because $L$ is positive semidefinite with $i(L) = [n-1 \ 0 \ 1]$ and $L_1$ is an $n-1$ order principal submatrix of $L$, according to the Cauchy's interlace theorem (Meyer, 2000), $L_1$ is positive definite. Thus, $\det(L_1) > 0$. Accordingly, we derive the following from (40), (41), and (42):

$$1_n^T Q 1_n = \sum_{i=1}^n k_i^2 (r_i + c_i) - \frac{v_{ref}^2}{P}\left(\sum_{i=1}^n k_i\right)^2 < 0 \quad (43)$$

The proof is accomplished.

**Remark**: As demonstrated by the proof, this theorem can solve this kind of inertia problem and obtain the critical condition. The condition of (28) can be regarded as the constraint of the system maximal load, which is defined as $P_{sup}$. Obviously, increasing the voltage can increase the upper bound of load. Thus, the condition of (28) is easy to be satisfied.

If (28) holds, the eigenvalues of matrix $J_2$ are all non-negative. However, what happens when a rank-one matrix $J_3$ is added? Additionally, how to design the voltage recovery coefficient $b_2$ such that the matrix $J_1$ is positive stable? In the following, the answers will be given by two theorems.

**Theorem 2.** If (28) is satisfied, the matrix $J_2$ will be diagonalizable.

**Proof.** First, the matrix $KLK$ is symmetric; thus, it is diagonalizable. Note

$$P_1 KLK P_1^T = \begin{bmatrix} \Lambda_1 & \\ & 0 \end{bmatrix}, J_4 = P_1 J_2 P_1^T,$$

$$\Lambda_1 = diag\left(\lambda_2(KLK), \cdots, \lambda_n(KLK)\right).$$

The matrix $P_1$ is an orthogonal matrix; thus, the matrix $J_4$ is isospectral with $J_2$. We reshape the matrix $J_3$ as follows

$$J_4 = P_1(C+Z)^{-1}KLKP_1^T = P_1(C+Z)^{-1}P_1^T\begin{bmatrix}\Lambda_1 & \\ & 0\end{bmatrix} \quad (44)$$

Note that $M = P_1(C+Z)^{-1}P_1^T$, and $M$ is symmetric. Thus, the matrix $J_4$ can be expressed as

$$J_4 = \begin{bmatrix} M_{11} & M_{12} \\ M_{12}^T & M_{22} \end{bmatrix}\begin{bmatrix}\Lambda_1 & \\ & 0\end{bmatrix} = \begin{bmatrix} M_{11}\Lambda_1 & 0_{n-1} \\ M_{12}^T\Lambda_1 & 0 \end{bmatrix} \quad (45)$$

where $M_{11}$, $M_{12}$, $M_{22}$ are the suitable dimensional submatrices of $M$. Because $\Lambda_1$ is a positive diagonal, and the matrix $M_{11}$ is symmetric, the matrix $M_{11}\Lambda_1$ is diagonalizable. That is, there is a nonsingular matrix $P_2$ such that $P_2 M_{11}\Lambda_1 P_2^{-1} = \Lambda_2$ and $\Lambda_2 = diag(\mu_2, \cdots, \mu_n)$. Note that $P_3 = P_2 \oplus 1$, $J_5 = P_3 J_4 P_3^{-1}$ and $P_4 = [I \ 0_{n-1} ; M_{12}^T P_2^{-1}\Lambda_2^{-1} \ 1]$. We have

$$P_4^{-1}J_5 P_4 = P_4^{-1}\begin{bmatrix}\Lambda_2 & 0_{n-1}\\ M_{12}^T P_2^{-1} & 0\end{bmatrix}P_4 = \begin{bmatrix} I & 0_{n-1} \\ -M_{12}^T P_2^{-1}\Lambda_2^{-1} & 1\end{bmatrix}\times$$
$$\begin{bmatrix}\Lambda_2 & 0_{n-1}\\ M_{12}^T P_2^{-1} & 0\end{bmatrix}\begin{bmatrix} I & 0_{n-1} \\ M_{12}^T P_2^{-1}\Lambda_2^{-1} & 1\end{bmatrix} = \begin{bmatrix}\Lambda_2 & \\ & 0\end{bmatrix} \quad (46)$$

According to (44) and (46), we have

$$P_5^{-1} J_2 P_5 = \Lambda_3 \quad (47)$$

where $P_5 = P_1^T P_3^{-1} P_4$ and $\Lambda_3 = \Lambda_2 \oplus 0$. Thus, we can determine whether the matrix $J_2$ is diagonalizable if and only if $\Lambda_2$ is nonsingular. The proof is accomplished.

Note $P_5 = [p_1 \ p_2 \ \cdots \ p_n]$, and define $P_6$ as

$$P_6 = P_5 T, T = diag\left\{\frac{1}{\|p_i\|_2}\right\} \quad (48)$$

Then, all column vectors of matrix $P_6$ are unitized.

**Remark:** Since (43) is always satisfied, matrix $J_2$ is diagonalizable. It will play a crucial role in the subsequent proof.

**Theorem 3**. If (28) is satisfied, matrix $J_1$ will be positive stable if the following conditions are satisfied.

$$b_1 > \gamma_1 b_2, \quad \gamma_1 = \sum_{i=1}^{n-1}\frac{f(\alpha_i\beta_i)}{\mu_{i+1}} \quad (49)$$

where $\beta = P_6^T 1_n$, $\alpha = r_L P_6^{-1}(C+Z)^{-1}g$ and the function f(x) is dedined as $f(x) = x$ if $x < 0$; otherwise, $f(x) = 0$.

**Proof.** According to **Theorem 2**, because (28) is satisfied, the matrix $J_2$ is diagonalizable. Note that $J_6 = P_6^{-1} J_1 P_6$; thus, $J_6$ is isospectral with $J_1$ and matrix $J_6$ can be expressed as

$$J_6 = b_1\Lambda_3 + b_2\alpha\beta^T \quad (50)$$

The characteristic equation of $J_6$ can be obtained as

$$|\lambda I - b_1\Lambda_3 - b_2\alpha\beta^T| = |\lambda I - b_1\Lambda_3||I - (\lambda I - b_1\Lambda_3)^{-1}b_2\alpha\beta^T|$$
$$= \lambda\left(\prod_{i=2}^n (\lambda - \mu_i)\right)\left(1 - \sum_{i=1}^{n-1}\frac{\alpha_i\beta_i}{\lambda - \mu_{i+1}} - \frac{\alpha_n\beta_n}{\lambda}\right) \quad (51)$$

We define the vector $\eta, \xi$ as $\eta_i = \alpha_i, \xi_i = \beta_i$ if $\alpha_i\beta_i \neq 0$ and as $\eta_i = \xi_i = 0$ if $\alpha_i\beta_i = 0$. Note that $J_7 = b_1\Lambda_3 + b_2\eta\xi^T$; according to (51), $J_7$ has the same characteristic equation as $J_6$. Thus, $J_7$ has the same eigenvalues as $J_6$.

Note that $J_8 = \Lambda_4 J_7 + J_7^T\Lambda_4, G_1 = \Lambda_4\eta\xi^T + \xi\eta^T\Lambda_4$, where $\Lambda_4$ is an undetermined positive definite diagonal matrix ($\Lambda_4 = diag\{\varepsilon_i\}$), then $J_8 = 2b_1\Lambda_3\Lambda_4 + b_2 G_1$. According to Lyapunov's stability theorem, if matrix $J_8$ is positive definite, $J_6$ is positive stable. According to the corollary presented in (Horn & Johnson, 2012), the eigenvalues of matrix $G_1$ can be expressed as

$$\lambda_1 = \eta^T\xi + \|\eta\|_2\|\xi\|_2, \lambda_2 = \eta^T\xi - \|\eta\|_2\|\xi\|_2, \lambda_3 = \cdots = \lambda_n = 0.$$

According to the Cauchy−Schwartz inequality, the matrix $G_1$ has at most one negative root. Thus, according to **Lemma 4**, it is easy to determine that the matrix $J_8$ has at most one negative eigenvalue. That is, the matrix $J_8$ is positive definite if and only if its determinant is positive. According to **Lemma 5**, the determinant of matrix $J_8$ can be obtained as

$$\det(J_8) = (2b_1)^{n-1}b_2\prod_{i=1}^{n-1}\mu_{i+1}\varepsilon_i \times$$
$$\left(2\eta_n\xi_n - \frac{1}{2}\sum_{i=1}^{n-1}\frac{b_2}{b_1\mu_{i+1}\varepsilon_i}(\eta_n\xi_i - \varepsilon_i\eta_i\xi_n)^2\right) \quad (52)$$

According to the foregoing analysis, matrix $J_1$ is positive stable if the following condition holds

$$2\eta_n\xi_n - \frac{1}{2}\sum_{i=1}^{n-1}\frac{b_2}{b_1\mu_{i+1}\varepsilon_i}(\eta_n\xi_i - \varepsilon_i\eta_i\xi_n)^2 > 0 \quad (53)$$

Because $b_1 \in (0, +\infty)$, (52) is solvable if and only if $\alpha_n\beta_n$ is positive. In the next part, we analyze the sign of $\alpha_n\beta_n$. According to **Lemma 5**,

$$\det(J_1) = \det(J_6) = b_2\alpha_n\beta_n b_1^{n-1}\prod_{i=1}^{n-1}\mu_{i+1} \quad (54)$$

By substituting (40), (41), and (42) into (54), we obtain

$$\det(J_1) = b_2(1_n^T Q1_n)\alpha_n\beta_n b_1^{n-1}\frac{\det(L_1)}{\det(Q)} \quad (55)$$

We define $x = K^{-1}g, y = r_L K^{-1}1_n$; then, the matrix $J_1$ can be equivalent to $J_1 = (C+Z)^{-1}K(b_1L + b_2xy^T)K$. Because $\det(Q) = \det(K^{-1})\det(C+Z)\det(K^{-1})$, we obtain

$$\det((C+Z)^{-1})\det(K)\det(L+xy^T)\det(K) =$$
$$b_2(1_n^T Q1_n)\alpha_n\beta_n b_1^{n-1}\frac{\det(L_1)}{\det(K^{-1})\det(C+Z)\det(K^{-1})} \quad (56)$$

By simplifying (56), $\alpha_n\beta_n$ is expressed as

$$\alpha_n\beta_n = \frac{\det(b_1L + b_2xy^T)}{b_2 b_1^{n-1}(1_n^T Q1_n)\det(L_1)} \quad (57)$$

Because $L$ is a balanced Laplacian matrix, there is an orthogonal matrix $P_7$ that satisfies $P_7^T LP_7 = \Lambda_5$, where $\Lambda_5 = diag\{\lambda_i(L)\}$ and $\lambda_n(L) = 0$. Note that $P_7 = [q_1\ q_2\ \cdots\ q_n]$, and $q_i$ is the corresponding unit eigenvector. For a connected graph, we obtain $\ker(L) = \text{span}(1_n)$, that is

$$p_n = \frac{1}{\sqrt{n}}1_n \quad (58)$$

Then, (57) can be transformed into

$$\alpha_n\beta_n = \frac{\det(b_1\Lambda_4 + b_2(P^T x)(P^T y)^T)}{b_2 b_1^{n-1}(1_n^T Q1_n)\det(L_1)} \quad (59)$$

According to the Matrix−tree Theorem (Molitierno, 2012), the absolute value of the determinant of any $(n-1)\times(n-1)$ submatrix of $L$ is equal to the number of spanning trees. Thus,

$$\det(L_1) = \frac{1}{n}\left(\prod_{i=1}^{n-1}\lambda_i(L)\right) \quad (60)$$

By substituting (60) in (59), the final analytical expression of $\alpha_n\beta_n$ is

$$\alpha_n\beta_n = \frac{(1_n^T x)(1_n^T y)}{1_n^T Q1_n} = \frac{r_L}{1_n^T Q1_n}\left(\sum_{i=1}^n k_i g_i\right)\left(\sum_{i=1}^n k_i\right) \quad (61)$$

Because $k_i$ and $g_i$ are non-negative, $r_L$ is negative, $1_n^T Q1_n$ is negative (if (28) is satisfied), and $\alpha_n\beta_n$ is positive.

Because $\alpha_n\beta_n$ is positive, (53) is equivalent to

$$b_1 > \frac{b_2}{4}\sum_{i=1}^{n-1}\frac{(\alpha_n\xi_i - \varepsilon_i\eta_i\beta_n)^2}{\alpha_n\beta_n\mu_{i+1}\varepsilon_i} \quad (62)$$

We set $\varepsilon_i = \sqrt{\left|\frac{\alpha_n\beta_i}{\beta_n\alpha_i}\right|}$ if $\alpha_i\beta_i \neq 0$ and $\varepsilon_i = 1$ otherwise. Thus, (49) is obtained. The proof is accomplished.

**Remark:** Equations (60) and (49) show that if $J_2$ is semi-positive and the determinant of $J_1$ is positive, $J_1$ is positive stable as long as $b_1$ is efficiently large. The matrices $J_2$ and $J_1$ can be regarded as a special case of the matrices $AB$ and $AB+F$, respectively, where $i(A) = [n-1\ 1\ 0]^T$ and $i(B) = [n-1\ 0\ 1]^T$, $F$ is a rank-one matrix. The techniques used in this study are effective for determining the positive stable conditions of the aforementioned matrices.

In summary, the system without time delay is stable if (28) and (49) are satisfied.

### 3.3. Stability conditions with communication delay

Time delays are often unavoidable in control systems based on communication and long time delay may be detrimental to stability. So, this section investigates the stability condition of the system with time delay. We assume that all the delays are the same, thus, the equivalent model is obtained from (26)-(27)

$$\frac{d\Delta i_i}{dt} = -J_1 \Delta i_i(t-\tau) \tag{63}$$

where $\tau$ is the time delay. The characteristic equation of the system is obtained as

$$\left|sI + J_1 e^{-\tau s}\right| = 0 \tag{64}$$

The system is stable if and only if all the roots of (64) are in the left half-plane. Define $\chi_1, \chi_2, \cdots, \chi_n$ as the eigenvalues of $J_1$, which can be expressed as $\chi_i = \Theta_i e^{j\theta_i}$, $(\Theta_i > 0, |\theta_i| < \pi)$.

**Theorem 4.** The roots of (64) are in the left half-plane if and only if

$$\tau\Theta_i + |\theta_i| < \frac{\pi}{2} \tag{65}$$

**Proof.** According to the well-known Schur's theorem about triangulation, there must exist a unitary matrix $U$ such that

$$U^H \left|sI + J_1 e^{-\tau s}\right| U = \left|sI + \Xi e^{-\tau s}\right| = 0 \tag{66}$$

where $\Xi$ is the similar upper triangular matrix. Then, (66) is decoupled as

$$s + \Theta_i e^{j\theta_i} e^{-\tau s} = 0 \tag{67}$$

Multiplying (67) by $\tau$ and then taking $z = \tau s$, we obtain

$$z + \tau\Theta_i e^{j\theta_i} e^{-z} = 0 \tag{68}$$

According to **Lemma 6**, (65) is obtained.

**Corollary 2.** A necessary condition of (65) is $\text{Re}(\chi_i) > 0$.

**Proof.** If there is an eigenvalue of $J_1$ in the left half-plane, we define it as $-p+qj$. Write it as the exponential form

$$-p+qj = \sqrt{p^2+q^2} e^{j(\pi-\delta)}, \quad \delta = \arccos\frac{p}{\sqrt{p^2+q^2}} \in \left(-\frac{\pi}{2}, \frac{\pi}{2}\right) \tag{69}$$

Then, (65) admits no solution because $|\pi-\delta| > \frac{\pi}{2}$. So, (65) admits solutions only when $\text{Re}(\chi_i) > 0$. The proof is accomplished.

In summary, the system with delay is stable if (28), (49) and (65) are satisfied.

**Remark:** In this section, the stability conditions of the system with delay are obtained. **Corollary 2** reveals that (28) and (49) are the necessaary conditions of (65). The relation among the eigenvalues of $J_1$ and delay that keep the system stable is obtained. Thus, the maximum delay bound that keep system stable is obtained.

## 4. Simulation

To verify the aforementioned analysis, simulations are performed in MATLAB/Simulink. The studied DC microgrid is shown in Fig.3. There are six DGs and one common CPL. All DGs are connected through communications. Three adjacent DGs (DG4, DG5, DG6) sample the load voltage and participate in the voltage recovery. The communication links are assumed bidirectional to feature a balanced Laplacian matrix. Each source is driven by a buck converter. All the buck converters supply a common CPL. The line resistances are $r_1 = r_2 = r_6 = 2$ Ω, $r_3 = 1$ Ω, and $r_4 = r_5 = 0.5$ Ω, and the reference voltage for the load is $v_{\text{ref}} = 200$ V.

### A. Control parameters design

The Laplacian matrix of the communication graph is designed as

$$L = \begin{bmatrix} 200 & -100 & & & & -100 \\ -100 & 200 & -100 & & & \\ & -100 & 200 & -100 & & \\ & & -100 & 200 & -100 & \\ & & & -100 & 200 & -100 \\ -100 & & & & -100 & 200 \end{bmatrix}$$

We assume that the proportion of the output current among the DGs is $2:2:2:1:1:1$. Then, we set $k_1 = k_2 = k_3 = 1$ and $k_4 = k_5 = k_6 = 2$. According to Fig.3, only DG4, DG5 and DG6 participate in the load voltage recovery, that is, $g_1 = g_2 = g_3 = 0$ and $g_4 = g_5 = g_6 = 1$. We set the virtual resistances as $c_1 = c_2 = c_3 = 5$ and $c_4 = c_5 = c_6 = 10$.

According to **Theorem 1** and **3**, the system without time delay is stable if (28) and (49) hold. According to (28), the maximum load of the system is obtained as $P_{\text{sup}} = 21.3$ kW. Define $\Delta_1$, and $\Delta_2$ as: $\Delta_1 = P - P_{\text{sup}}, \Delta_2 = b_1 - \gamma_1 b_2$.

If $\Delta_1$ and $\Delta_2$ are all positive, the system will be stable.

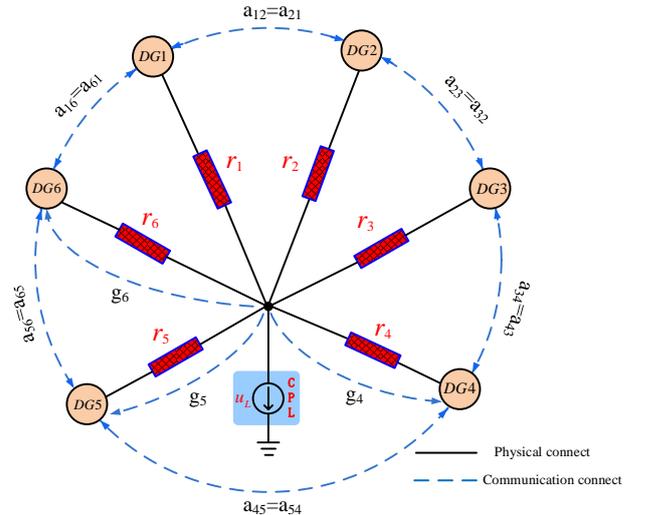

Fig.3. Simulated prototype of a DC microgrid.

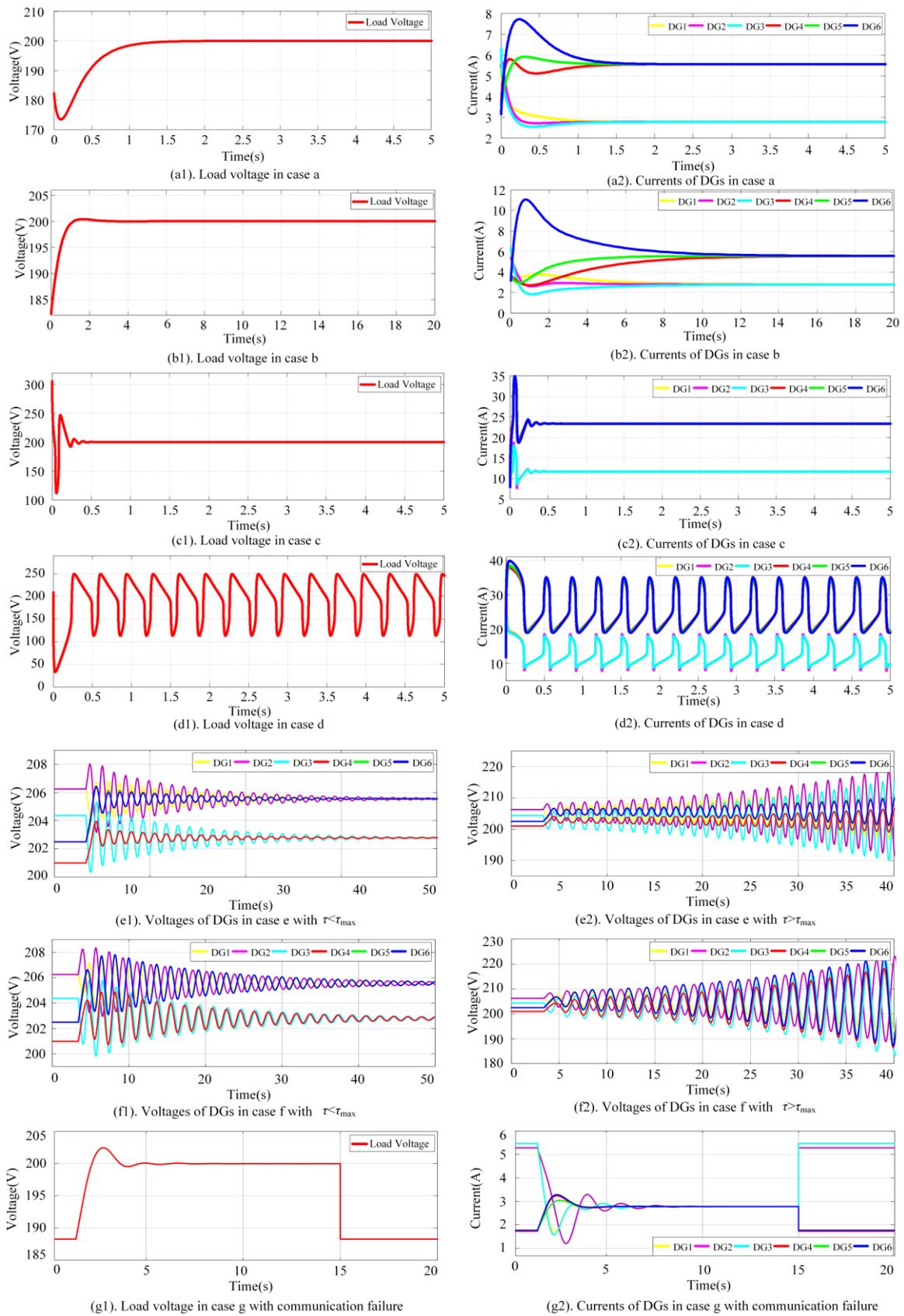

Fig.4. Simulation results.

### B. Steady-state and transient performances of the system without delay

In order to test the correctness of the stability conditions, four cases are tested:

a: $P = 5$ kW, $\gamma_1 = 0.06$, $b_1 = 20$, $b_2 = 10$, $\Delta_1 > 0, \Delta_2 > 0$;
b: $P = 5$ kW, $\gamma_1 = 0.06$, $b_1 = 1$, $b_2 = 20$, $\Delta_1 > 0, \Delta_2 < 0$;
c: $P = 21$ kW, $\gamma_1 = 114.4$, $b_1 = 300$, $b_2 = 2$, $\Delta_1 > 0, \Delta_2 > 0$;
d: $P = 21.4$ kW, $b_1 = 20$, $b_2 = 10$, $\Delta_1 < 0$.

It is worthy emphasizing that (49) is derived when (28) is satisfied; thus, we only calculate $\Delta_1$ for case d. The current sharing, voltage recovery and stability are tested for these cases simultaneously.

Fig. 4. (a1), (b1), (c1), (a2), (b2), and (c2) show that the load voltage is 200V, and the proportion of the current among the DGs is 2 : 2 : 2 : 1 : 1 : 1; thus, the current sharing and voltage recovery can be realized simultaneously if the system is stable. Fig. (a1) and (c1) show that if (28) and (49) hold, the system is stable. Fig (c1) and (d1) show that case c is stable and case d is unstable, respectively, which means that (28) is necessary and the critical value of the load power is accurate. In case b, (49) is not satisfied, but the system is still stable, which reveals that (49) is a sufficient condition for the system stability. To summarize, the simulation results coincide with the theoretical analysis.

### C. The system with delay and link failures

Due to the possible time delay and link failures, introducing the communication will reduce the reliability. In this section, we test the stability of the system with time delay and link failures. According to (65), the high gain may lead to the system instability when the system has time delay. Take the Laplacian matrix of the communication graph $L_1=0.01L$.

**Time-delay:** According to **Theorem 3** and **corollary 2**, if (28), (49) and (65) are satisfied, the system will stable. Define $\tau_{\max} = \min\left\{\dfrac{\pi - 2|\theta_i|}{2\Theta_i}\right\}$ as the maximal delay that keeps the system stable. In order to test the correctness of the stability conditions, six cases are tested:

Case e1: $P$=5kW, $b_1$=4, $b_2$=5, $\tau_{\max}$=0.3, $\tau$=0.29;
Case e2: $P$=5kW, $b_1$=4, $b_2$=5, $\tau_{\max}$=0.3, $\tau$=0.31;
Case f1: $P$=5kW, $b_1$=3, $b_2$=6, $\tau_{\max}$=0.4, $\tau$=0.39;
Case f2: $P$=5kW, $b_1$=3, $b_2$=6, $\tau_{\max}$=0.4, $\tau$=0.41;

In case e1, e2, f1 and f2, the system works in droop control mode before t=3s, then distributed control is introduced.

Fig. 4 (e1), (e2), (f1) and (f2) reveal that the system is stable when $\tau > \tau_{\max}$, otherwise, it becomes unstable.

**Link failures:** in order to investigate the influence of the possible communication failures on the performances of the proposed method, case g is tested.
Case g: $P$=5kW, $b_1$=5, $b_2$=4, $\tau$=0.4;

The case g is carried out according to the following sequence of actions:

1) Starting the system with only traditional droop control;
2) Applying distributed control at t=1s;
3) The link 1-6 and 4-load fail ($a_{16}=a_{61}=0$, $g_4=0$) at t=5s;
4) All the communication links fail at t=15s.

In Fig. 4 (g1) and (g2), it shows that the traditional droop control leads to voltage deviation and cannot achieve accurate power sharing before t=1s. In contrast, after introducing distributed control, despite partial communication links failures at t=5s, accurate power sharing and voltage regulation are achieved, for the reason that there still exists a spanning tree. At t=15s, even if all the communication links break down, the proposed control becomes traditional droop control.

### 5. Conclusions

A distributed control that aims at current sharing, voltage recovery and overcoming the CPL instability is proposed. The small-signal stability of the system with time delay is analyzed. The stability conditions are obtained strictly by theoretical deduction. The main contribution of this paper is that the relation among the line resistances, control parameters, reference voltage and the maximum load that keep the system stable is obtained. In this sense, the analytical conditions offer a design guideline to build reliable microgrids. Additionally, this paper has provided an effective method to analyze the inertia of two typical matrices. The current sharing, voltage recovery and stability of the system under the proposed method are verified through simulations. However, the proposed method can not achieve global asymptotically stable. For future research, we will further study the conservativeness issue in the condition and the control synthesis for global stbilization.

### Appendix. Proof of Lemma 5.

**Lemma** 5. Define $\Lambda = c_1 \oplus c_2 \cdots \oplus c_{n-1} \oplus 0$, $M_1 = \Lambda + ab^T$, $M_2 = M_1 + ba^T$, $a = [a_1\ a_2\ \cdots\ a_n]^T$, and $b = [b_1\ b_2\ \cdots\ b_n]^T$. Then,

$$(1)\ \det(M_1) = a_n b_n \prod_{i=1}^{n-1} c_i$$

$$(2)\ \det(M_2) = \prod_{i=1}^{n-1} c_i \left(2 a_n b_n - \sum_{i=1}^{n-1} \frac{1}{c_i}(a_n b_i - a_i b_n)^2\right)$$

**Proof.** We define $\Gamma = c_1 \oplus c_2 \cdots \oplus c_{n-1} \oplus x$, where $x$ is an infinitesimal variable. Then, we obtain the determinant of $M_1$ as follows

$$\det(M_1) = \lim_{x \to 0} \det(\Gamma + ab^T)$$
$$= \lim_{x \to 0} \det(\Gamma)\det(I + \Gamma^{-1} ab^T) \qquad (70)$$

Becaus $rank(\Gamma^{-1}ab^T) = 1$, $\Gamma^{-1}ab^T(\Gamma^{-1}a) = (b^T\Gamma^{-1}a)(\Gamma^{-1}a)$, the matrix $\Gamma^{-1}ab^T$ has the unique nonzero eigenvalue $b^T\Gamma^{-1}a$, together with $n-1$ zeros. Thus, the following is easily obtained

$$\det\left(I+\Gamma^{-1}ab^T\right)=1+\sum_{i=1}^{n-1}\frac{a_ib_i}{c_i}+\frac{a_nb_n}{x} \quad (71)$$

By substituting (71) into (70), we obtain

$$\det(M_1)=\lim_{x\to 0}\left(x\prod_{i=1}^{n-1}c_i\right)\left(1+\sum_{i=1}^{n-1}\frac{a_ib_i}{c_i}+\frac{a_nb_n}{x}\right) \quad (72)$$
$$=a_nb_n\prod_{i=1}^{n-1}c_i$$

Similarly, we obtain the determinant of the matrix $M_2$ as follows:

$$\det(M_2)=\lim_{x\to 0}\det\left(\Gamma+ab^T+ba^T\right)$$
$$=\lim_{x\to 0}\det(\Gamma)\det\left(I+\Gamma^{-1}ab^T+\Gamma^{-1}ba^T\right) \quad (73)$$

We define $W=\Gamma^{-1}ab^T+\Gamma^{-1}ba^T$; thus, the matrix $W$ can be expressed as $W=\begin{bmatrix}\Gamma^{-1}a & \Gamma^{-1}b\end{bmatrix}\begin{bmatrix}b & a\end{bmatrix}^T$. We define the matrix $W_1$ as $\begin{bmatrix}b & a\end{bmatrix}^T\begin{bmatrix}\Gamma^{-1}a & \Gamma^{-1}b\end{bmatrix}$. Then, the matrix $W$ has the same nonzero eigenvalues as $W_1$, and we easily obtain that $\lambda_{1,2}(W_1)=b^T\Gamma^{-1}a\pm\sqrt{(a^T\Gamma^{-1}a)(b^T\Gamma^{-1}b)}$. The matrix $W$ has the eigenvalues $\lambda_{1,2}(W_1)$, together with $n$-2 zeros.

Then, we obtain

$$\det\left(I+\Gamma^{-1}ab^T+\Gamma^{-1}ba^T\right)=\left(1+b^T\Gamma^{-1}a+\sqrt{(a^T\Gamma^{-1}a)(b^T\Gamma^{-1}b)}\right)$$
$$\times\left(1+b^T\Gamma^{-1}a-\sqrt{(a^T\Gamma^{-1}a)(b^T\Gamma^{-1}b)}\right)$$
$$=1+2b^T\Gamma^{-1}a+\left(b^T\Gamma^{-1}a\right)^2-\left(a^T\Gamma^{-1}a\right)\left(b^T\Gamma^{-1}b\right) \quad (74)$$

By substituting (74) into (73), the following is obtained

$$\det(M_2)=\lim_{x\to 0}x\prod_{i=0}^{n-1}c_i\left(\left(1+\sum_{i=1}^{n-1}\frac{a_ib_i}{c_i}\right)^2-\left(\sum_{i=1}^{n-1}\frac{a_i^2}{c_i}\right)\left(\sum_{i=1}^{n-1}\frac{b_i^2}{c_i}\right)\right)$$
$$+\lim_{x\to 0}x\prod_{i=0}^{n-1}c_i\left(\frac{2a_nb_n}{x}+\frac{2a_nb_n}{x}\sum_{i=1}^{n-1}\frac{a_ib_i}{c_i}-\frac{a_n^2}{x}\left(\sum_{i=1}^{n-1}\frac{b_i^2}{c_i}\right)-\frac{b_n^2}{x}\left(\sum_{i=1}^{n-1}\frac{a_i^2}{c_i}\right)\right)$$
$$=\prod_{i=1}^{n-1}c_i\left(2a_nb_n-\sum_{i=1}^{n-1}\frac{1}{c_i}(a_nb_i-a_ib_n)^2\right) \quad (75)$$

Thus, the proof is accomplished.

## Acknowledgements

This work is supported by the National Natural Science Foundation of China under Grants 51677195, 61622311, 61573384, Fundamental Research Funds for the Central Universities of Central South University under Grant 2016zzts052, and the Project of Innovation-driven Plan in Central South University.

## References

Simpson-Porco, J.W., Dörfler, F., & Bullo, F. (2013). Synchroniza-tion and power sharing for droop-controlled inverters in islanded microgrids. *Automatica*, 49, 2603-2611.

George, C. K., Zhong, Q., Ren, B., & Krstic, M. (2015). Bounded droop controller for parallel operation of inverters. *Automatica*, 53, 320-328.

Schiffer, J., Zonetti, D., Ortega, R., Stanković, A. M., & TevfikSezi, J. R. (2016). A survey on modeling of microgrids—From fundamental physics tophasors and voltage sources. *Automatica*, 74, 135-150.

Bidram1, A., Lewis, F. L., & Davoudi, A. (2014).Synchronization of nonlinear heterogeneous cooperative systems using input-output feedback linearization. *Automatica*, 50, 2578-2585.

Song, D., Yang J., Cai Z., Dong M., Su M & Wang, Y. (2017). Wind estimation with a non-standard extended Kalman filter and its application on maximum power extraction for variable speed wind turbines. *Applied Energy*, 190, 670-685.

Schiffer, J., Zonetti, D., Ortega, R., Stanković, A. M., & TevfikSezi, J. R. (2014). Conditions for stability of droop-controlled inverter-based microgrids. *Automatica*, 50, 2457-2469.

Chang, C., & Zhang, W. (2016). Distributed control of inverter-based lossy microgrids for power sharing and frequency regulation under voltage constraints. *Automatica*, 66, 85-95.

Sun, Y., Hou, X., Yang, J., Han, H., Su, M., & Guerrero, J. M. (2017). New perspectives on droop control in AC microgrid. *IEEE Trans. Ind. Electronics.* 64(7), 5741-5745.

Kakigano, H., Miura, Y., & Ise, T. (2010). Low-voltage bipolar-type DC microgrid for super high quality distribution. *IEEE Trans. Power Electron.*. 25(12), 3066-3075.

Maknouninejad, A., Qu, Z. Lewis, F. L. & Davoudi, A. (2014). Optimal, nonlinear, and distributed designs of droop controls for DC microgrids. *IEEE Trans. Smart Grid*. 5(5), 2508–2516.

Han, H., Wang, H., Sun, Y., Yang, J., & Liu, Z. (2017). A distributed control scheme on cost optimization under communication delays for DC microgrids. *IET Generation Transmission & Distribution*.to be published.

Augustine, S., Mishra, M. K., & Lakshminarasamma, N. (2015). Adaptive droop control strategy for load sharing and circulating current minimization in low-voltage standalone DC microgrid. *IEEE Trans. Sustainable energy*. 6(1), 132-141.

Huang, P., Liu, P., Xiao, W., & Moursi, M. S. E. (2015). A novel droop-based average voltage sharing control strategy for DC microgrids. *IEEE Trans. Smart Grid*. 6(3), 1096–1106.

Guo, L., Feng, Y., Li, X., & Wang, C. (2014). Stability analysis of a DC microgrid with master-slave control structure. In *Proc. IEEE Energy Conversion Congress and Exposition,* Pittsburgh, PA, 5682-5689.

Nasirian, V., Moayedi, S., Davoudi, A., & Lewis, F. L. (2015). Distributed cooperative control of DC microgrids. *IEEE trans. Industry Applications,* 30(4), 2288-2303.

Nasirian, V., Davoudi, A., Lewis, F. L., & Guerrero, J. M. (2014). Distributed adaptive droop control for DC distribution systems. *IEEE Trans. Energy Conversion,* 29(4), 944-956.

Moayedi, S., Student, G., & Davoudi, A. (2016). Distributed tertiary control of DC microgrid clusters. *IEEE Trans. Power Electron.* 31(4), 1717-1733.

Shafiee, Q., Dragicevic, T., Vasquez, J. C., & Guerrero, J. M. (2014). Hierarchical control for multiple DC-microgrids clusters. *IEEE Trans. Energy Conversion.* 29(4), 922-933.

Zhao, J., & Dörfler, F. (2015). Distributed control and optimization in DC microgrids. *Automatica.* 61, 18-26.

Behjati, H., Davoudi, A., & Lewis, F. (2014). Modular DC-DC converters on graphs: cooperative control. *IEEE* Trans. *Power Electron*, 29(12), 6725-6741.

Meng, L., Dragicevic, T., Roldán-Pérez, J., Vasquez, J. C., & Guerrero, J. M. (2016). Modeling and sensitivity study of consensus algorithm-based distributed hierarchical control for DC microgrids. IEEE Trans. Smart Grid. 7(3), 1504-1515.

Sandeep, A., & Fernandes B. G. (2013). Reduced-order model and stability analysis of low-voltage DC microgrid. *IEEE Trans. Industrial Electron.*.60(11), 5040-5049.

Tahim, A. P. N., Pagano, D. J., Lenz, E., & Stramosk, V. (2015). Modeling and stability analysis of islanded DC microgrids under droop control. *IEEE Trans. Power Electron.*.30(8), 4597-4607.

Su, M., Liu, Z., Sun, Y., Han, H., & Hou, X. (2016). Stability analysis and stabilization methods of DC microgrid with multiple parallel-connected DC-DC converters loaded by CPLs. *IEEE Trans. Smart Grid.*to be published.

Tisseur, F., & Meerbergen, K. (2001). The quadratic eigenvalue problem, SIAM Rev. 43, 235–286.

Olfati-Saber, R., & Murray, R. M. (2004).Consensus problems in networks of agents with switching topology and time-delays. *IEEE Trans. Automatic Control.* 49(9), 1520-1533.

Horn, R. A., & Johnson, C. R. (1986). Topics in matrix analysis. Cambridge.

Ostrowski, A., & Schneider, H. (1962). Some theorems on the inertia of general matrices. *Journal of Mathematical analysis and applications*, 4, 72-84.

Hong, Y., & Horn, R. A. (1991). The Jordan canonical form of a Hermitian and a positive semidefinite matrix. *Linear Algebra and Its Application*, 147, 373-386.

Horn, R. A., & Johnson, C. R. (2012). Matrix analysis. Cambridge university press.

Meyer, C. D. (2000). Matrix analysis and applied linear algebra. society for industrial and applied mathematics. Philadelphia.

Molitierno, J. J. (2012).Applications of combinatorial matrix theory to Laplacian matrices of graphs. CRC Press.

DeFazio, M. V. & Muldoon, Martin E (2005). On the zeros of a transcendental function. *Advances In Analysis*, 385-394.

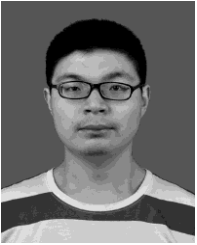

**Zhangjie Liu** received the B.S. degree in Detection Guidance and Control Techniques from the Central South University, Changsha, China, in 2013, where he is currently working toward Ph.D. degree in control engineering.

His research interests include Renewable Energy Systems, Distributed generation and DC micro-grid.

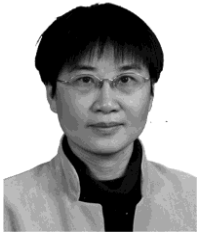

**Mei Su** was born in Hunan, China, in 1967. She received the B.S., M.S. and Ph.D. degrees from the School of Information Science and Engineering, Central South University, Changsha, China, in 1989, 1992 and 2005, respectively.

Since 2006, she has been a Professor with the School of Information Science and Engineering, Central South University. Her research interests include matrix converter, adjustable speed drives, and wind energy conversion system.

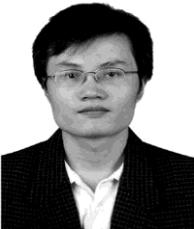

**Yao Sun** (M'13) was born in Hunan, China, in 1981. He received the B.S., M.S. and Ph.D. degrees from the School of Information Science and Engineering, Central South University, Changsha, China, in 2004, 2007 and 2010, respectively. He is currently with the School of Information Science and Engineering, Central South University, China, as an associate professor.

His research interests include matrix converter, micro-grid and wind energy conversion system.

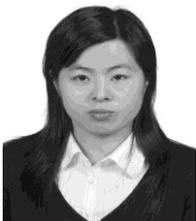

**Hua Han** was was born in Hunan, China, in 1970. She received the M.S. and Ph.D. degrees from the School of Information Science and Engineering, Central South University, Changsha, China, in 1998 and 2008, respectively. She was a visiting scholar of University of Central. Florida, Orlando, FL, USA, from April 2011 to April 2012. She is currently a professor with the School of Information Science and Engineering, Central South University, China.

Her research interests include microgrid, renewable energy power generation system and power electronic equipment.

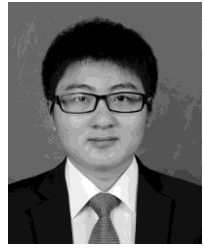

**Xiaochao Hou** received the B.S. degree in Automation and M.S. degrees in Control Science and Engineering from the Central South University, Changsha, China, in 2014 and 2017, respectively, where he is currently working toward Ph.D. degree in electrical engineering.

His research interests include Renewable Energy Systems, Distributed generation and micro-grid.

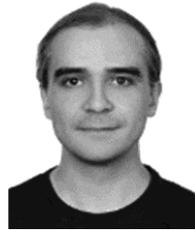

**Josep M. Guerrero** received the B.S. degree in telecommunications engineering, the M.S. degree in electronics engineering, and the Ph.D. degree in power electronics from the Technical University of Catalonia, Barcelona, in 1997, 2000 and 2003, respectively.

Since 2011, he has been a Full Professor with the Department of Energy Technology, Aalborg University, Denmark, where he is responsible for the Microgrid Research Program. His research interests is oriented to different microgrid aspects, including power electronics, distributed energy-storage systems, hierarchical and cooperative control, energy management systems, and optimization of microgrids and islanded minigrids.